\documentclass[aps,prstab,reprint,superscriptaddress,floatfix]{revtex4-1}
\usepackage{graphicx,amsmath}

\begin{document}


\title{High throughput data streaming of individual longitudinal electron bunch profiles in a storage ring with single-shot electro-optical sampling}



\author{Stefan Funkner}
\email{stefan.funkner@kit.edu}
\affiliation{Laboratory for Applications of Synchrotron Radiation, Karlsruhe Institute of Technology, Kaiserstra{\ss}e 12, D-76131, Germany}
\author{Edmund Blomley}
\affiliation{Institute for Beam Physics and Technology, Karlsruhe Institute of Technology, Hermann-von-Helmholtz-Platz 1, D-76344 Eggenstein-Leopoldshafen, Germany}
\author{Erik Br\"{u}ndermann}
\affiliation{Institute for Beam Physics and Technology, Karlsruhe Institute of Technology, Hermann-von-Helmholtz-Platz 1, D-76344 Eggenstein-Leopoldshafen, Germany}
\author{Michele Caselle}
\affiliation{Institute for Data Processing and Electronics, Karlsruhe Institute of Technology, Hermann-von-Helmholtz-Platz 1, D-76344 Eggenstein-Leopoldshafen, Germany}
\author{Nicole Hiller}
\altaffiliation[Present address:]{PSI, Villigen, Switzerland}
\affiliation{Laboratory for Applications of Synchrotron Radiation, Karlsruhe Institute of Technology, Kaiserstra{\ss}e 12, D-76131, Germany}
\author{Michael J. Nasse}
\affiliation{Institute for Beam Physics and Technology, Karlsruhe Institute of Technology, Hermann-von-Helmholtz-Platz 1, D-76344 Eggenstein-Leopoldshafen, Germany}
\author{Gudrun Niehues}
\affiliation{Institute for Beam Physics and Technology, Karlsruhe Institute of Technology, Hermann-von-Helmholtz-Platz 1, D-76344 Eggenstein-Leopoldshafen, Germany}
\author{Lorenzo Rota}
\affiliation{Institute for Data Processing and Electronics, Karlsruhe Institute of Technology, Hermann-von-Helmholtz-Platz 1, D-76344 Eggenstein-Leopoldshafen, Germany}
\author{Patrik Sch\"{o}nfeldt}
\affiliation{Institute for Beam Physics and Technology, Karlsruhe Institute of Technology, Hermann-von-Helmholtz-Platz 1, D-76344 Eggenstein-Leopoldshafen, Germany}
\author{Sophie Walther}
\altaffiliation[Present address:]{DESY, Hamburg, Germany}
\affiliation{Laboratory for Applications of Synchrotron Radiation, Karlsruhe Institute of Technology, Kaiserstra{\ss}e 12, D-76131, Germany}
\author{Marc Weber}
\affiliation{Institute for Data Processing and Electronics, Karlsruhe Institute of Technology, Hermann-von-Helmholtz-Platz 1, D-76344 Eggenstein-Leopoldshafen, Germany}
\author{Anke-Susanne M\"{u}ller}
\affiliation{Laboratory for Applications of Synchrotron Radiation, Karlsruhe Institute of Technology, Kaiserstra{\ss}e 12, D-76131, Germany}
\affiliation{Institute for Beam Physics and Technology, Karlsruhe Institute of Technology, Hermann-von-Helmholtz-Platz 1, D-76344 Eggenstein-Leopoldshafen, Germany}



\date{\today}

\begin{abstract}
The development of fast detection methods for comprehensive monitoring of electron bunches is a prerequisite to gain comprehensive control over the synchrontron emission in storage rings with their MHz repetition rate. Here, we present a proof-of-principle experiment with at detailed description of our implementation to detect the longitudinal electron bunch profiles via single-shot, near-field electro-optical sampling at the Karlsruhe Research Accelerator (KARA). Our experiment is equipped with an ultra-fast line array camera providing a high-throughput MHz data stream. We characterize statistical properties of the obtained data set and give a detailed description for the data processing as well as for the calculation of the charge density profiles, which where measured in the short-bunch operation mode of KARA. Finally, we discuss properties of the bunch profile dynamics on a coarse-grained level on the example of the well-known synchrotron oscillation.     
\end{abstract}


\maketitle

\section{INTRODUCTION}
\begin{figure*}
\includegraphics[scale=0.5]{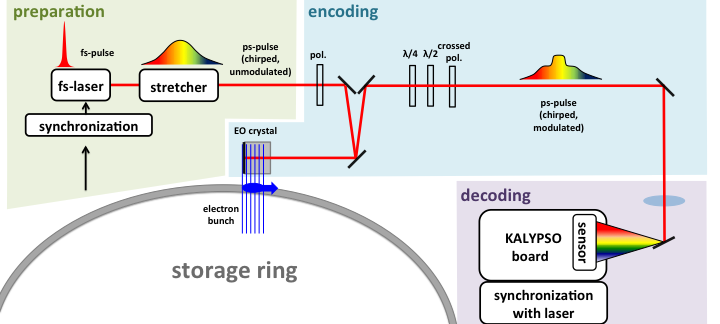}
\caption{\label{setup} The setup to measure longitudinal charge density profiles of electron bunches at KARA is shown: The radiation is prepared by providing linearly chirped laser pulses synchronized to the bunch repetition rate, then the bunch profile is encoded onto the laser pulse in an electro-optical crystal and finally, the decoding of the bunch profile is performed using an ultra-fast line array (KALYPSO).}
\end{figure*}
To experimentally verify theoretical predictions on complex bunch dynamics in storage ring, the challenge is given by the typically high MHz-range repetition rates in combination with the requirement of a non-destructive single-shot technique to detect dynamics without averaging. \\
The phase space of electron bunches in a storage ring is subject to a continuous long-term evolution. This is of particular importance, if coherent synchrotron light is produced by relativistic electron bunches with their lengths compressed to the millimeter or even sub-millimeter range. In such a short bunch mode, the ensemble of electrons in a single bunch interacts with its own radiation field every time the bunch is deflected in the bending magnets. This self-interaction can lead to instabilities in the long-term dynamics of the electron bunch when the number of electrons exceed a certain threshold. As a result, microstructures can form spontaneously in the phase space \cite{Szwaj2016}. Because these microstructures produce strong THz radiation bursts with an intensity more than 5 orders of magnitude higher than incoherent synchrotron radiation \cite{Evain2017} potentially leading to e.g. applications in non-linear spectroscopy, this process is subject to intense investigations \cite{wustefeld2010coherent,Abo-Bakr2003,Shields2012,warnock2000general,Stupakov2002,Bane2010,Evain2012,Schonfeldt2017,Caselle:2014mza,Brosi2016}.\\  
Although far-field CSR measurements are a useful approach, additional methods exploring the phase space more directly are desirable in order to tighter connect theory and experiment. Here, an option is to use EOS techniques to sense the near-field of the electron bunches and provide direct information about the phase space. EOS is a well-developed, convenient technique to detect the electrical field of THz transients when using laser-based THz spectroscopy. Fast detection rates for EOS can be achieved with single-shot methods \cite{Jiang1998}. \\
In 2002, the EOS technique was adapted by Wilke et. al to measure the near-field of electron bunches in a linear accelerator at the FELIX free electron laser facility \cite{Wilke2002}. By introducing single-shot EOS, Wilke et. al measured the length of individual relativistic electron bunches. In 2013, the first experimental realization in a storage ring was demonstrated by our group by mounting an electro-optical crystal inside the vacuum chamber of the Karlsruhe Research Accelerator (KARA) in close vicinity of the electron bunches \cite{Hiller2013a,Hiller2011,Hiller2013}. \\
While single-shot detection of bunch profiles could be demonstrated with these experiments, the low repetition rate of the commercial DAQ system of a few Hz made it impossible to study the short bunch beam dynamics. Therefore, DAQ systems with higher repetition rates up to the revolution frequency are needed. Subsequently, the so-called photonic time-stretch method, which measures the temporal profile of electron bunches with repetition rates in the MHz range, has been demonstrated for far-field measurements at SOLEIL \cite{Roussel2015} and in a combined setup with near-field measurements at KARA \cite{Szwaj2017}.\\
In this publication, we present a sampling method based on an ultra-fast detector line array. This technique permits gap-less ultra-fast data streaming and therefore makes in-depth studies of the bunch profile dynamics occurring at various time scales over long observation times possible. With the present publication, we aim to provide a comprehensive description of our experimental setup and data evaluation. In order to provide an exemplary application, we discuss basic features of the coherent synchrotron oscillation detected with our method.  \\
This paper is organized as follows: First, we provide a detailed description of the experimental implementation and data processing required to obtain the electron bunch profile. In the results section, we present data measured at 0.91 MHz repetition rate during the microbunching instability of an electron bunch, and give a brief characterization of the main properties of the datasets. While we concentrate the discussion on the coarse-grain scale of measured charge densities, recent preliminary results obtained with a higher repetition rate corresponding to the revolution frequency of KARA and an increased signal-to-noise ratio, which show substructures on the charge densities due to microbunching, can be found in \cite{Rota2016}.  Finally, we will give a short summary and outlook to future investigations and improvements.

\section{MATERIALS AND METHODS}

The experiments, outlined in this section, were performed at the storage ring KARA at KIT. During the measurements, the storage ring was operated at a beam energy of 1.3 GeV, with a radio frequency (RF) voltage of 1500 kV and a bunch current of 0.35 mA stored in a single bunch (for additional parameters see \cite{Brosi2016}). The storage ring was operated in the so-called low-$\alpha_c$ mode \cite{Muller2005} compressing the electron bunch length to a few picoseconds. \\
Our setup, displayed in Figure \ref{setup}, is based on well-known techniques for single-shot electro-optical sampling measurement of THz field transients \cite{Jiang1998}. The operation principle relies primarily on three interconnected methods, namely preparation, encoding and decoding. Below, we describe our adaption of these methods to measure charge density profiles of electron bunches at KARA (see in particular \cite{Hiller2013a}). 

\subsection{Preparation: Providing chirped laser pulses synchronized to the bunch repetition rate}

To measure the charge density (or longitudinal bunch profile) ${\rho}(t)$ of electron bunches, we use coherent laser pulses as a spectral ruler for the time dependence by implementing a linear time-frequency correlation $t({\omega})$. \\
Our starting point is a custom-built laser \cite{Hiller2013a} based on an Ytterbium-doped fiber oscillator with a repetition rate of 62.5 MHz and a central wavelength of 1030 nm. A critical point for the measurements is the synchronization of the laser to the revolution frequency of the storage ring. Here, an active synchronization is realized and locked to the 500 MHz RF master oscillator. A pulse picker is used to reduce the laser repetition rate, and a single-pass fiber amplifier outputs pulses with a central wavelength of about 1050 nm, a spectral width of about 80 nm, and a typical optical power in the range of a few mW \cite{Hiller2013a}. The measurements are performed in single-bunch operation at a repetition rate of 2.72 MHz. However, the front-end electronics of our data acquisition system, at the time of the measurements reported in this paper, is limited to 1 Mfps. Thus, the laser pulse picker is set to a repetition rate of about 0.91 MHz enabling a detection of every third revolution of the single electron bunch. \\
The 0.91 MHz pulse train is sent via an optical fiber from the laboratory outside the shielding walls to the storage ring. The intrinsic dispersion of the optical fiber stretches the fs-pulse to several ps. Afterwards, the laser pulse passes a grating compressor, which allows the adjustment of the final pulse duration to overlap the laser pulse with the complete electron bunch. By changing the length of the short pulses at the laser output (in the ideal case they are Fourier-transform-limited) with the optical fiber and the compressor afterwards, their instantaneous frequency ${\omega}$ becomes chirped so that a linear time-frequency correlation $t({\omega})$ is established. 

\subsection{Encoding the bunch profile onto the laser pulses}

While the instantaneous frequency represents the time axis, the polarization of the light encodes the charge density of the electron bunch (see \cite{Steffen2007, Casalbuoni2008} for a detailed description). Therefore, the laser pulses are transmitted through a polarizer to ensure a high degree of linear polarization. Then they are sent in an in-vacuum setup in the storage ring through an electro-optical, gallium phosphide (GaP) crystal placed close to the electron beam. A highly reflecting back-side of the crystal sends the pulses back (see Figure 1). After this reflection, the laser pulses travel parallel with the electron bunch. As the electron bunch is traveling at highly relativistic speed with $\gamma = 2544$ \cite{Hiller2013a}, the electric field of the electron bunch is squeezed perpendicular to the direction of propagation (in the laboratory reference frame). Therefore, assuming that the bunch can be modeled as a line charge density, the temporal profile of the electrical near-field leaking into the GaP crystal resembles the charge density profile ${\rho}(t)$ of the electron bunch. These time-dependent fields propagate through the crystal changing its birefringence due to the Pockels effect. As a consequence, the phase retardation is modified according to:
\begin{equation}
\Gamma(t) = \frac{2\pi d}{\lambda_0}n_0^3 r_{41} E(t) \text{ ,}
\end{equation}
where $r_{41}$ is the Pockels coefficient, $d$ is the crystal thickness, $\lambda_0$ the average laser wavelength in vacuum, $n_0$ the index of refraction at the laser wavelength and $E(t)  \propto {\rho}(t)$  the electrical field \cite{Casalbuoni2008} (thus, $\Gamma(t)$ might be interpreted as  $\Gamma(\rho(t))$). The latter leads via $t = t({\omega})$ to a frequency-dependent encoding of the bunch profile onto the polarization state of the chirped laser pulse. \\
In the next step, the laser pulses exit the storage ring and pass a $\lambda/4$ waveplate to compensate for the static intrinsic birefringence of the EO crystal, and a $\lambda/2$ waveplate to adjust the polarization of an unmodulated pulse to be nearly crossed (at an angle $\theta$ of $4.6^\circ$) with respect to the subsequent polarizer. This adjustment is a compromise between linearity of the response to the electric field and a good signal-to-noise ratio.\\
After the polarizer, the electron bunch profile is spectrally encoded onto the frequency-dependent intensity of the broadband laser pulses according to:
\begin{equation}
I_{\text{M}}({\omega},{\rho})= \frac{1}{2} I_{\text{laser}}({\omega})\cdot (1-\cos({\Gamma}({\rho})-4{\theta})) \text{ ,}
\end{equation}
where $I_{\text{M}}$ is the intensity of the modulated laser pulse after the polarization optics and $I_{\text{laser}}$ is the intensity spectrum of the laser in front of the polarizer \cite{Hiller2013a}. Let $I_{\text{U}}(\omega )=I_{\text{M}} (\omega ,0)$ be the unmodulated signal when no electron bunch is present. Defining the relative signal modulation as $S_{\text{mod}}= I_{\text{M}} (\omega, \rho )/I_{\text{U}} (\omega)$ and using $\cos (\Gamma (\rho )- 4 \theta) \approx \cos (4 \theta) +\Gamma \sin (4\theta)$, we can approximate $S_{\text{mod}}$ by
\begin{equation}
S_{\text{mod}} =  \frac{I_{\text{M}}}{I_{\text{U}}} \approx 1+ \frac{\sin (4\theta)}{1-\cos (4\theta)}  \Gamma (\rho ).
\end{equation}
Hence, the wavelength-dependent quantity $S_{\text{mod}}-1$ is proportional to the bunch profile. 

\subsection{Decoding the bunch profile with a high repetition data acquisition scheme}
To decode this information, we sent the laser pulses via an optical fiber from the storage ring to an optical table in the EO laboratory, where we use a grating to disperse every pulse into its spectral components. We focus the pulses with a lens onto the 256-pixel silicon line array of the KIT-developed ultra-fast spectrometer ``{\bf KA}rlsruhe {\bf L}inear arra{\bf Y} detector for MHz-re{\bf P}etition rate {\bf S}pectr{\bf O}scopy'' (KALYPSO)\cite{Rota2014} measuring every third revolution of the electron bunch in a single-shot scheme. Previous calibration measurements showed that we obtain a linear dependence of the time axis (at a ps-scale) and the pixel number. In the next section, we describe the processing of the raw data to recover the bunch profiles. 

\subsection{Data processing}
\begin{figure*}
\includegraphics[width=0.9\linewidth]{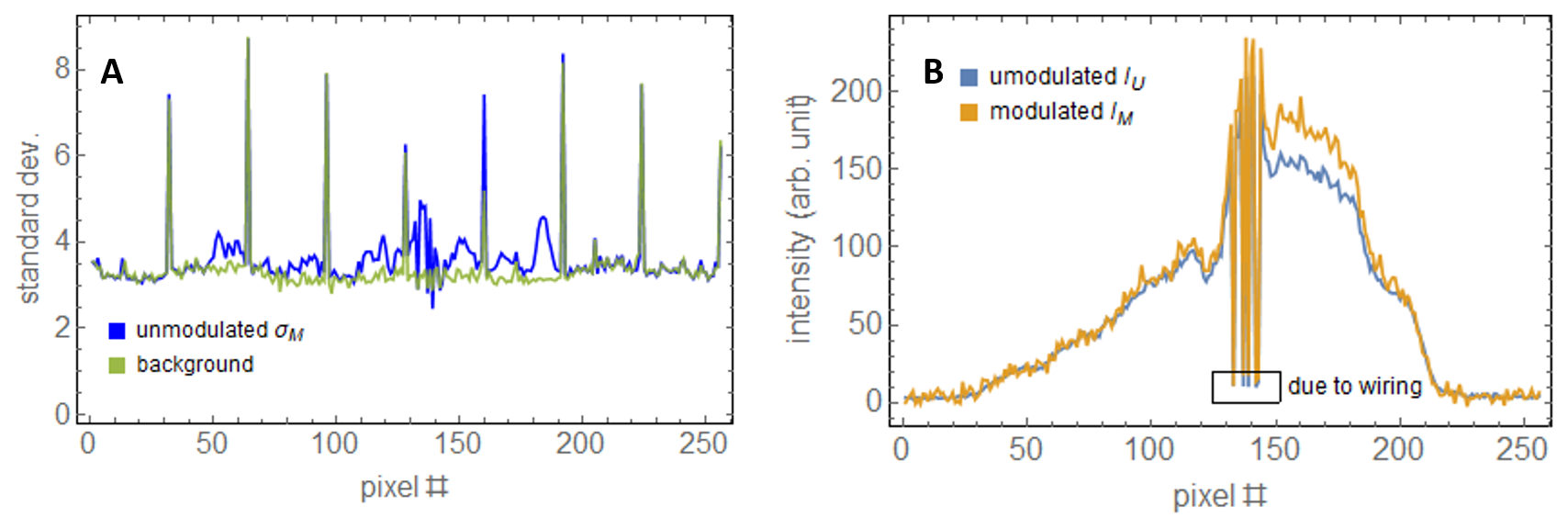}
\caption{\label{rawdata} A) Standard deviation of a background signal (obtained with a blocked laser beam) and an unmodulated signal. We evaluated 100,000 measurements for both plots.  B) Averaged intensity of the same 100,000 measurements of the unmodulated laser pulse from panel A compared with a single-shot measurement of a modulated laser pulse. Both data sets were corrected by subtracting the background measured with a blocked laser beam.}
\end{figure*}

\begin{figure*}
\includegraphics[width=0.9 \linewidth]{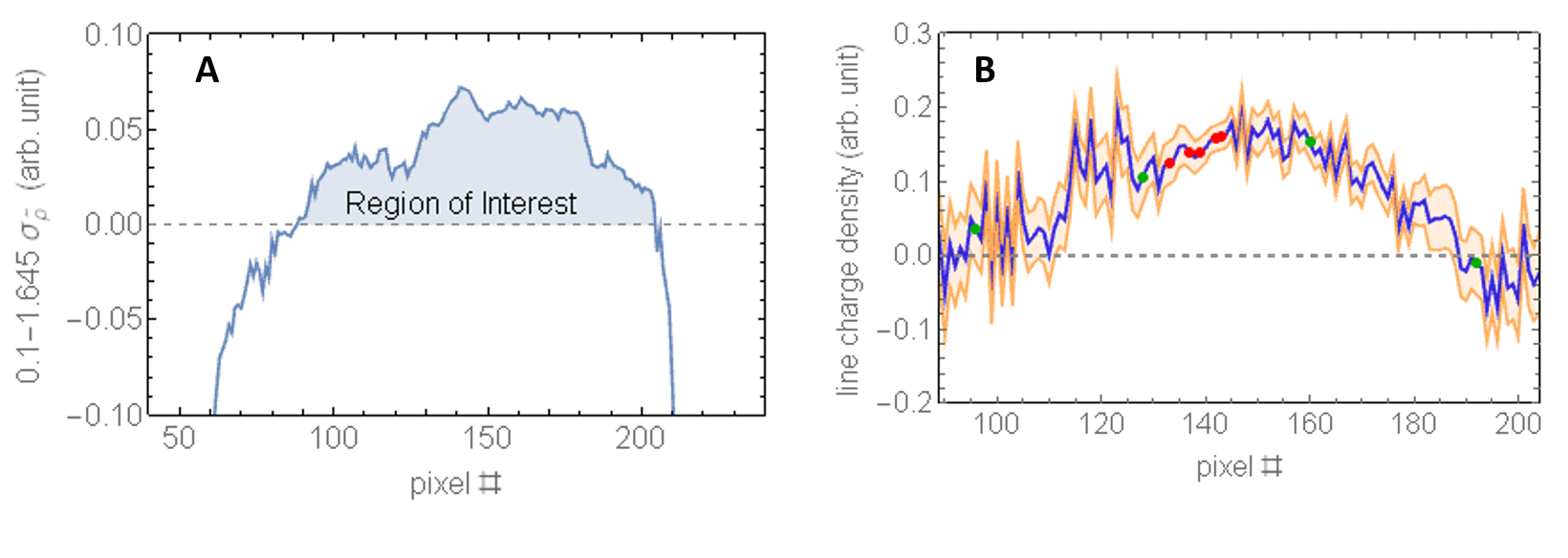}
\caption{\label{singleCP} Panel A) displays a visualization of the selection criterion defining a pixel region of interest, considered for further data evaluation. B) An example of an electron bunch density reconstruction for a single-shot measurement (blue line). The red and green dots indicate linearly interpolated data points from pixels with spikes originated from insufficient wire bonding and timing issues, respectively. The latter show up as spikes in the standard deviation of the modulated signal. The orange bands indicate the 68.3\% confidence interval for the value of the reconstructed line charge density.}
\end{figure*}

In order to retrieve the longitudinal bunch profiles, we need to account for detector effects like the background signal or other intrinsic sensor properties from the KALYPSO electronics as well as stray light. To do this, we measured 100,000 shots with the laser beam blocked and subtracted the average from every single-shot measurement. The standard deviation of the background signal, shown in Figure \ref{rawdata}A, is relatively constant for all pixels besides some spikes visible for every 32nd pixel. We attribute these spikes to a sub-optimal configuration of the preliminary KALYPSO firmware. A comparison with the standard deviation of an unmodulated signal in Figure \ref{rawdata}A demonstrates that the uncertainty in our intensity measurements is dominated by background noise of detector system.\\
A comparison between a typical single-shot measurement of a modulated signal and a 100,000 shot average of the unmodulated signal is displayed in Figure \ref{rawdata}B. In the range between pixel 150 and 200, the intensity modulation due to the electron bunch (max. 20 \%) is clearly visible. Furthermore, sharp spikes in the range of pixel 133 to 143 were present in both measurements. These spikes are due to defective wire-bonding connections between the Si sensor and the readout electronics of the used KALYPSO prototype board.  \\
From the previous statements, we assume for the bunch profile $\rho_i \propto I_{\text{M},i}/I_{\text{U},i} -1 =: \tilde{\rho}_i $, where we introduced for convenience $\tilde{\rho}$ as a dimensionless quantity proportional to $\rho$. Here, $i$ is the pixel number or rather the time index. As $I_{\text{U}}$ is estimated from the 100,000 shot average, the uncertainty of $\tilde{\rho}$ is dominated by the uncertainty of the single-shot measurement $I_{\text{M}}$. Calculating the propagation of uncertainty, we get $\sigma_{\tilde{\rho},i} \approx  \sigma_{\text{M},i}/ I_{\text{U},i}$ , where $\sigma_{\text{M},i}$ would be the  standard deviation from hypothetical repeated single-shot measurements under the exact same conditions (in the experiment the electron bunch evolves so that the same conditions cannot be reproduced exactly again). In Figure \ref{rawdata}A we show that the standard deviation stays nearly constant independent of whether the background or the unmodulated signal is measured. As the change in intensity between the modulated and unmodulated signal is smaller than between the unmodulated signal and the background, it is reasonable to assume that $\sigma_{\text{M},i} \approx \sigma_{\text{U},i}$. Hence, we conclude that 
\begin{equation}
\sigma_{\tilde{\rho},i}  \approx  \sigma_{\text{U},i}/ I_{\text{U},i} \text{ .}
\end{equation}
Figure \ref{rawdata}A also demonstrates that $\sigma_{\text{U},i}$ is nearly independent from the pixel number $i$ so that the shape of $\sigma_{\tilde{\rho},i}$ is dominated by the term $1/I_{\text{U},i}$ . As the reference signal $I_{\text{U},i}$, shown in Figure \ref{rawdata}B, decays rapidly at the edges of the measured range, the standard deviation $\sigma_{\tilde{\rho}}$ increases leading to the question which pixels should be considered for further data evaluation. Here, we decided for the following selection criterion:
We consider all pixels where half of the maximum of the signal modulation $S_{\text{mod}}$  (in our case 0.1 for an assumed maximum of the signal modulation of 0.2) can be distinguished from the background (i.e. no modulation) with an accuracy of 90 \%, which corresponds to a factor of 1.645 when assuming a Student's-t distribution. Hence, all pixel numbers $i$ for which
\begin{equation}
0.1-1.645 \,\sigma_{\tilde{\rho},i}>0
\end{equation}
is true (see Figure \ref{singleCP}A) are selected for further data evaluation. While the choice of the criterion can still be subject of some bias, its concrete formulation helps to compare our findings to other experiments. Figure \ref{singleCP}A displays the quantity 0.1-1.645 $\sigma_{\tilde{\rho},i}$. The region of interest selected by the criterion includes the pixel numbers between 89 and 204.\\ 
Finally, we calculated the line charge density of the electron bunch $\rho_i \propto \tilde{\rho}_i$. An example result of such a calculation is displayed in Figure \ref{singleCP}B, where we also show in orange the 68.3\% confidence interval corresponding to $\pm\sigma_{\tilde{\rho},i}$. Here and in all subsequent considerations, we removed pixels with wire bonding or readout problems from our data set. For the replacement, we use a linear interpolation determined from the direct neighboring pixels to conserve the line shape. In order to analyze the results of a measurement series, we constructed revolution plots from stacked series of consecutive density profile measurements, which are described in the next section.    

\section{RESULTS}
\begin{figure*}
\includegraphics[width=0.93\linewidth]{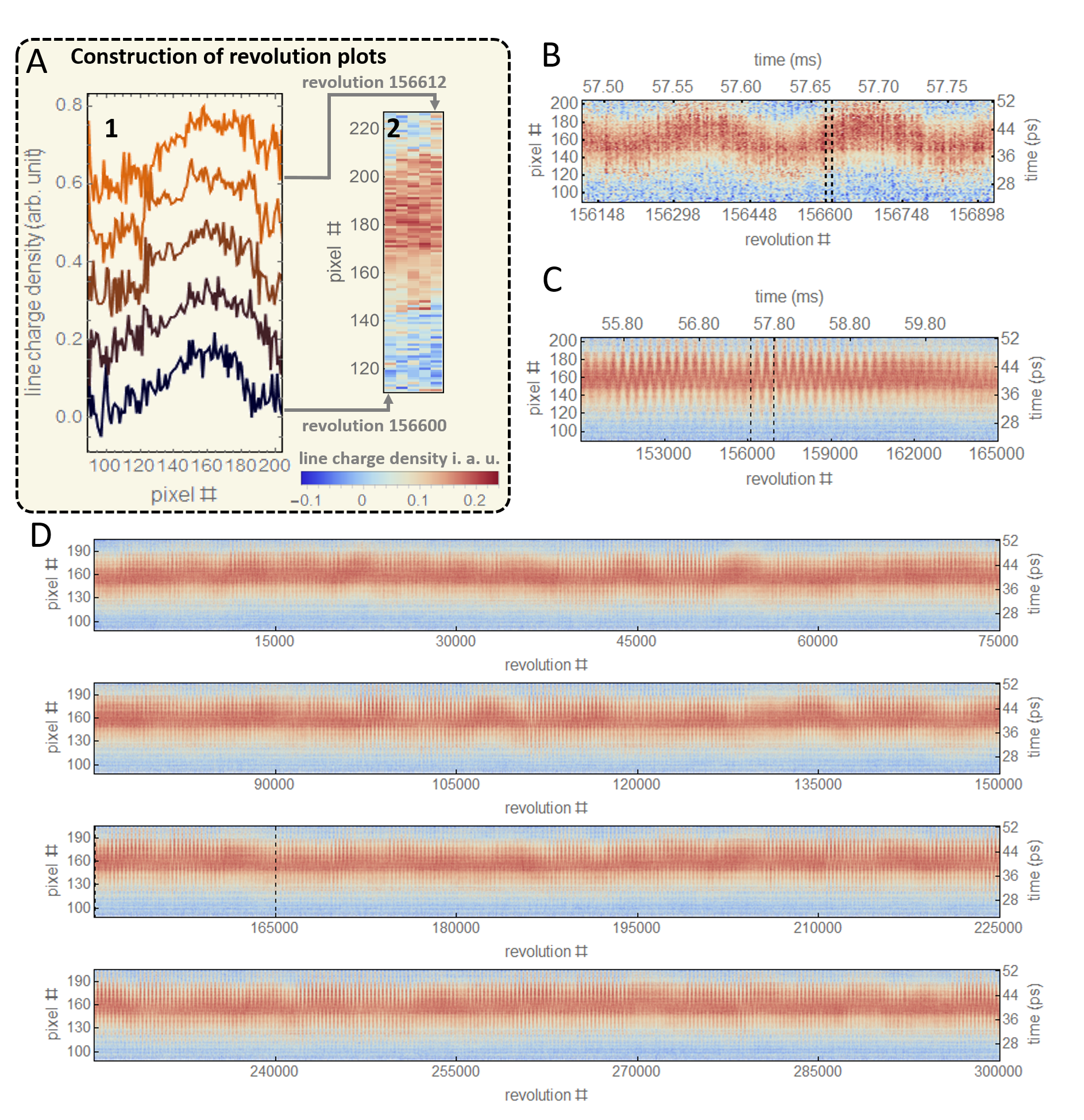}
\caption{\label{results} Presented is the construction of revolution plots from single-shot measurements in A) followed by B)-D) step-wise zoom outs with an increasing number of data points.
A) On the left side in part 1, 5 consecutive measurements are displayed via vertical stacking. On the left side in part 2, the corresponding revolution plot is shown. For the revolution plot, the measurements are stacked horizontally. Therefore, values of the line charged density are encoded by the color scheme shown at the bottom, for every pixel respectively. B) displays a time interval of 0.3 ms, C) a time interval of 5.55 ms and D) a time interval of 110.11 ms spanning the complete data stream of 100,000 measurements. The dashed lines refer to the time interval of the previous plot, respectively. The right time axis for each plot is obtained from previous calibration measurements by using the laser synchronization system. We estimate the conversion factor to be ($0.256 \pm 0.016 $) ps/pixel.}
\end{figure*}
\begin{figure*}
\includegraphics[width=0.9\linewidth]{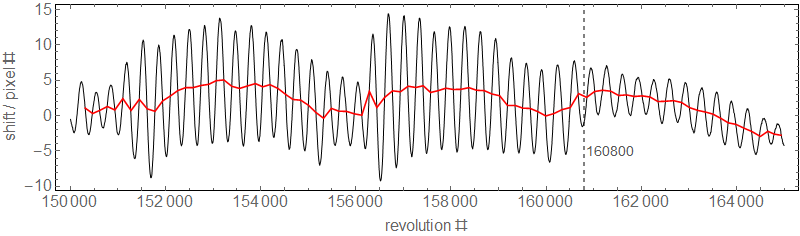}
\caption{\label{syncroOsz}Shift of the bunch profile center position in units of pixel numbers relative to the mean center position. To reduce the noise, we averaged over the 15 nearest neighbors (x-axis) for each data point (black line). The red line visualizes long-term drifts and is obtained by averaging the positions of consecutive maxima and minima.}
\end{figure*} 
Figure \ref{results}A visualizes the construction of a revolution plot. Our starting point are subsequent measurements of the line charge density. Figure \ref{results}A1 displays 5 stacked consecutive measurements. In the plot, which is constructed on the right side (Figure \ref{results}A2), each line corresponds to a single-shot measurement, while the color encodes the magnitude of the line charge density. Consecutive measurements are stacked from left to right to visualize the time-dependent dynamics of the bunch profile. For ease of reference this visualization is referred to as ``revolution plot''. \\ 
The data representation as a revolution plot simplifies the comparison between time frames and allows to easily identify time-dependent patterns in the bunch dynamics - at various zoom-levels or time ranges. For example, the revolution plot of the 5 consecutive measurements (Figure \ref{results}A2) might already indicate that the center of bunch profiles slowly shifts to higher pixel numbers, which is difficult to realize from Figure \ref{results}A1. In Figure \ref{results}B, we increased the number of stacked measurements. The 5 measurements from Figure \ref{results}A are displayed between the dashed lines revealing that the slow center shift of the bunches is part of an oscillation. By averaging the Fourier transform of the signal of each pixel (not shown here, for all measured data) we estimate that this frequency is ($8.343\pm 0.09$) kHz. This matches well with the synchrotron oscillation frequency of the electron bunch from the storage ring bunch-by-bunch feedback system \cite{Blomley:2016ttu}, which was found to be ($8.3\pm 0.1$) kHz.  We will discuss this feature in more detail at the end of this section.\\
The oscillation uncovers a measurement artifact: for pixel numbers below 144 the values of the line charge density are systematically decreased for all times frames (which can be seen as a slight step in the color coding). We attribute this effect to the insufficient wiring (in the range of pixel 133 to 143) mentioned in the methods section (see Figure \ref{rawdata}B). During our data processing we removed pixels with a wiring defect and estimated the missing data points by linearly interpolating adjacent pixels. The measured artifact might originate from an influence of the these pixels on their neighbors resulting in a non-linear response of these pixels to the intensity of the incident laser light. Thus, if the pixels become insensitive for increased light intensity, the light modulation and therefore the electron density is systematically underestimated.\\ 
Changes of the oscillation amplitude are visualized by a further zoom out in Figure \ref{results}C. The amplitude of the oscillation changes continuously during the observed time frame of a few milliseconds. Finally, a data stream of 100,000 consecutive measurements is displayed in Figure \ref{results}D corresponding to a time interval of 110.11 milliseconds. In particular, the large time intervals of Figures \ref{results}C and \ref{results}D reveal the complex dynamic behavior of the synchrotron oscillation in the low-alpha mode. \\
To take a closer look on the longitudinal motion of the bunch, we roughly estimate the center of mass by fitting a scaled normal distribution to each profile measurement in Figure \ref{results}C. As a result of the fits, we obtain the time-dependent center of mass position shown in Figure \ref{syncroOsz} in units of pixel numbers. As stated above, the frequency of this motion is the synchrotron frequency. \\
This oscillation, which is induced by the phase focusing of the electron bunch to the RF voltage of the storage ring, is a result of the rotation of the electrons in the longitudinal 2D phase space \cite{Wille2001} assuming that the distribution dynamics can be covered by linear approximations, while ignoring damping and self-interaction of the electron bunch. In this regard, we recognize the turn-by-turn changes of the electron density of states as a mapping according to a 2D rotation of the phase space coordinates \cite{Schonfeldt2017}. With respect to the phase space, an EO measurement is the projection of the electron density distribution on the phase space time axis. \\
The fact that the EO measurements can detect the synchrotron oscillation leads to the conclusion that the center of mass of density distribution is not exactly mapped on itself under the phase space rotation for consecutive time steps. As a consequence, the EO measurements detect an oscillation of the electron bunch with an amplitude depending on the distance between the rotation center and the center of mass (in the phase space). For example, the significant amplitude decrease above revolution 160800 might be related to a shift of the density distribution in phase space. \\   
The excitation of the center-of-mass synchrotron oscillation may have different causes. For example, the oscillation might be driven by amplitude and/or phase noise of the accelerating RF field \cite{Ormond}. Another cause might be the (current-dependent) emission of coherent synchrotron radiation due to the above described microbunching instability leading to a repeating increase in energy loss of the electrons. As a consequence, the equilibrium condition between energy gain in the RF field and energy loss due to the circular motion in the storage ring is disturbed leading to different phase focusing conditions and the excitation of the synchrotron motion \cite{Wille2001}. To get a deeper understanding of the electron bunch dynamics, a validation of the different contributions is desirable, but beyond the scope of this publication. \\ 
In Figure \ref{syncroOsz}, long-term drifts of the baseline are indicated by the red line, calculated from the mean of consecutive maxima and minima. The observed long-term drifts (on a few ms scale) of the center position might come from changes of the accelerator conditions or the phase space dynamics of the electron bunch.  Also, drifts in synchronization of the laser system to the master clock of the storage ring might be the cause of small baseline changes. As for the synchrotron oscillation more investigations are required.

\section{SUMMARY AND OUTLOOK}
We demonstrated, in a proof-of-principle experiment, measurements of the near-field of a compressed relativistic electron bunch in a storage ring by single-shot EO sampling showing the potential of the method for studies of longitudinal beam dynamics. \\
The detected signals are dominated by the synchrotron oscillation. With a repetition rate of 0.91 MHz, we are able to detect every third revolution. Therefore, the sampling rate of the presented setup is similar to that of photonic time-stretch experiments. By measuring 100,000 frames we demonstrated that the data stream can be continuously recorded. Here, the only limitation is given by the memory capacity. To overcome this limitation, we are currently implementing online data processing  algorithms based on Graphical Processing Units (GPUs) \cite{Vogelgesang2016}. This will enable real-time extraction of relevant bunch properties from the large amount of raw data produced by the KALYPSO detector, thus enabling continuous monitoring of the beam behavior.\\
With the Si photodiode array, we could well resolve the entire profile of the electron bunch and track unambiguously its synchrotron oscillation. In contrast to beam position monitors (BPMs), we are not only able to detect the arrival of the electron bunch at ps time scales, but also determine the profile. Therefore, the EO sampling technique bears the potential to distinguish between signal changes due to a change in arrival time or a change in the bunch profile, which is not provided by the BPM resolution. \\
Nonetheless, to resolve the micro-structuring of electron bunches responsible for the bursts of CSR in the THz range in short-bunch operation, an increased signal-to-noise ratio is required. This can be achieved by improving the signal-to-noise ratio of the KALYPSO detector.  Moreover, by implementing a balanced detection scheme similar to \cite{Evain2012}, the noise contribution introduced by single-shot fluctuations of the laser spectrum could be removed from the measurements. Currently, we are commissioning a new version of KALYPSO, with a frame-rate of 2.72 MHz and a better noise performance, enabling the detection of every single bunch revolution at KARA. Here, first preliminary result show that the microbunching of the electron bunches can be well resolved with our system \cite{Rota2016}. Finally, a KALYPSO version with an InGaAs sensor is foreseen for setups where the wavelength of the laser spectrum exceeds 1050 nm.

\begin{acknowledgments}
This work is funded by the BMBF contract numbers 05K13VKA and 05K16VKA. We acknowledge M. Brosi, B. Kehrer and J. L. Steinmann for many fruitful discussions.  
\end{acknowledgments}
\newpage

\bibliography{EOSD}
\end{document}